\documentclass[a4paper]{jpconf}
\usepackage{graphicx}

\begin{document}
\title{Black holes from generalized gauge field theories}

\author{J. Diaz-Alonso, D. Rubiera-Garcia}

\address{LUTH, Observatoire de Paris, CNRS, Universit\'e Paris
Diderot. 5 Place Jules Janssen, 92190 Meudon, France}
\address{Departamento de Fisica, Universidad de Oviedo. Avda.
Calvo Sotelo 18, E-33007 Oviedo, Asturias, Spain}

\ead{joaquin.diaz@obspm.fr ; diego.rubiera-garcia@obspm.fr}

\begin{abstract}
We summarize the main results of a broad analysis on electrostatic, spherically symmetric (ESS) solutions of a class of non-linear electrodynamics models minimally coupled to gravitation. Such models are defined as arbitrary functions of the two quadratic field invariants, constrained by several physical admissibility requirements, and split into different families according to the behaviour of these lagrangian density functions in vacuum and on the boundary of their domains of definition. Depending on these behaviours the flat-space energy of the ESS field can be finite or divergent. For each model we qualitatively study the structure of its associated gravitational configurations, which can be asymptotically \emph{Schwarzschild-like} or with an anomalous \emph{non Schwarzschild-like} behaviour at $r \rightarrow \infty$ (but being asymptotically flat and well behaved anyhow). The extension of these results to the non-abelian case is also briefly considered.
\end{abstract}

\section{Introduction}

The non-linear Born-Infeld (BI) model \cite{BI}, introduced in 1934 as an attempt to classically provide a finite energy for the electron field, has been the cornerstone of many developments in theoretical physics in the last three decades. The BI lagrangian density takes the form

\begin{equation}
L_{BI}=\beta^2 \left(1-\sqrt{1-\frac{X}{\beta^2}-\frac{Y^2}{4\beta^4}}\right), \label{BI}
\end{equation}
where $X=-\frac{1}{2}F_{\mu\nu}F^{\mu\nu}=\vec{E}^2-\vec{H}^2$ and $Y=-\frac{1}{2}F_{\mu\nu}F^{*\mu\nu}=2\vec{E} \cdot \vec{H}$ are the two quadratic field invariants that can be constructed with the field strength tensor $F_{\mu\nu}=\partial_{\mu}A_{\nu}-\partial_{\nu}A_{\mu}$ and its dual $F^{*\mu\nu}=\frac{1}{2}\epsilon^{\mu\nu\alpha\beta}F_{\alpha\beta}$, and where $\vec{E}$ and $\vec{H}$ are the electric and magnetic fields, respectively.

When minimally coupled to gravity, its electrostatic spherically symmetric (ESS) solutions have been studied for a long time \cite{gravity-BI}. With no doubt, the finding that string theory leads to (abelian and non-abelian) BI-like lagrangians in its low-energy limit \cite{string-BI}, has greatly contributed to the renewed interest in this kind of non-linear actions. As a result many non-linear electrodynamics (NED) models, aside from BI one, leading to asymptotically flat ESS structures, have been considered in the literature in the last decades \cite{NED}. Moreover, the characterization of the full class of physically admissible NED theories leading to asymptotically Schwarzschild-like configurations when coupled to gravity was performed in Ref.\cite{dr09}. Other models have been found to lead to well behaved gravitational configurations but approaching asymptotic flatness slower than the Schwarzschild term \cite{hassaine08,dr10}. These lines of research have also lead to some promising attempts in the obtention of regular (singularity-free) ESS black hole solutions in NEDs minimally coupled to gravity \cite{regular}. Let us finally mention that black hole solutions generated by NEDs in other gravitational backgrounds have been also considered \cite{NED-backgrounds}.

In this work we summarize the main results of a general procedure for the classification and proper characterization of asymptotically flat electrically charged black hole solutions of NEDs defined as arbitrary functions of the two quadratic field invariants \emph{without explicitly fixing the lagrangian density function}. The key point in such an analysis is the existence of a first-integral of the field equations, which takes the same form with and without gravity. These models are constrained by a number of physical \emph{admissibility} requirements and classified according to the asymptotic and central-field behaviours of their ESS solutions. Their associated flat-space energy can be finite or divergent, due to one or both of these field behaviours in the limits of the integral of energy ($r \rightarrow \infty$ and $r \sim 0$). For each one of these models we qualitatively characterize their associated (asymptotically flat and well behaved) gravitational configurations. Finally, the extension of these procedures to the non-abelian case is briefly discussed.

\section{Gravitating ESS fields}

Let us begin by introducing the action for NED models minimally coupled to the Einstein action as

\begin{equation}
S=\int d^4 x \sqrt{-g}\left(\frac{R}{16\pi G} - \varphi(X,Y)\right),
\end{equation}
where $\varphi(X,Y)$ is an arbitrary function of the two field invariants previously introduced. We consider an ansantz for the static spherically symmetric metric as

\begin{equation}
ds^2=e^{\nu(r)}dt^2-e^{\mu(r)}dr^2-r^2 d \Omega^2, \label{metric}
\end{equation}
where $d \Omega^2=d\theta^2+\sin^2 \theta d\phi^2$ is the interval on the $2$-sphere. However, due to the symmetry of the NED energy-momentum tensor $T_t^t=T_r^r$, only a single independent metric function remains as $\lambda(r)=e^{\nu(r)}=e^{-\mu(r)}$. The Einstein equations $G_{\mu\nu}=8\pi T_{\mu\nu}$ (we are taking units $G=1$) for this metric can be easily obtained to give \cite{dr09}

\begin{equation}
\lambda(r)=1-\frac{2M}{r}+\frac{2\varepsilon_{ex}(r,q)}{r}, \label{metric1}
\end{equation}
where $M$ is an integration constant, identified as the Arnowitt-Deser-Misner (ADM) mass, and we have introduced, by convenience, the \emph{exterior integral of energy}, defined as

\begin{equation}
\varepsilon_{ex}(r,q)=\int_r^{\infty} R^2 T_0^0(R,q)dR. \label{ex}
\end{equation}
However, the procedure leading to Eq.(\ref{metric1}) needs the implicit assumption that the NED source is such that $\varepsilon_{ex}(r,q)$ converges in its upper limit ($r \rightarrow \infty$). As we shall see in the sequel there are NED models not satisfying this property, and in such cases the metric is integrated as

\begin{equation}
\lambda(r)=1+\frac{C}{r}-\frac{2\varepsilon_{in}(r,q)}{r}, \label{metric2}
\end{equation}
where now $C$ is an integration constant not longer identified as the ADM mass (since, as we shall see, the large-$r$ behaviour is governed in this case by the electrostatic field term, rather than the ``Schwarzschild" one) and now

\begin{equation}
\varepsilon_{in}(r,q)=\int_0^{r} R^2 T_0^0(R,q)dR \label{in}
\end{equation}
is the \emph{interior integral of energy}. In this case, the obtention of Eq.(\ref{metric2}) needs the assumption that $\varepsilon_{in}(r,q)$ converges in the lower limit ($r \rightarrow 0$). But even if neither $\varepsilon_{ex}(r,q)$ nor $\varepsilon_{in}(r,q)$ can be defined, then the Einstein equations can be still integrated as

\begin{equation}
\lambda(r,q,D)=1+\frac{D}{r}-\frac{2\varepsilon(r,q,0)}{r}, \label{metric3}
\end{equation}
where

\begin{equation}
\varepsilon(r,q,B)=4\pi \int r^2T_0^0(r,q)+B, \label{epsilon3}
\end{equation}
$B$ and $D=C-2B$ being integration constants. Consequently, in order to properly characterize the metric function in the different cases, one is lead to analyze the conditions under which $\varepsilon_{ex}(r,q)$ and $\varepsilon_{in}(r,q)$ are well defined. This amounts to study the behaviour of $E(r)$ at $r \rightarrow \infty$ and around $r \sim 0$. To do this we take into account the fact that the NED field equations (we are defining $\varphi_X=\frac{\partial \varphi}{\partial X}$), obtained as $\nabla_{\mu}(\varphi_X F^{\mu\nu}+\varphi_Y F^{*\mu\nu})=0$ lead, for ESS fields ($\vec{E}(r)=E(r) \frac{\vec{r}}{r} \neq 0, \vec{H}(r)=0$), to a first-integral

\begin{equation}
r^2 \varphi_X E(r)=q, \label{FI}
\end{equation}
where $q$ is an integration constant identified as the electric charge associated to the field $E(r)$. Remarkably, this first-integral takes the same form, in the Schwarzschild coordinate system, as in the absence of gravity in spherical coordinates. Consequently, we can perform the classification of the models in flat space according to the asymptotic and central-field behaviours of $E(r)$, and Eq.(\ref{FI}) will allow the translation of the results obtained there to the gravitational context.

\subsection{Definition of the models}

Let us now establish a set of constraints on the NED models, leading to what shall be called \emph{admissible} models. We restrict the lagrangian density function $\varphi(X,Y=0)$ to be \emph{continuous, derivable and single-branched} in its domain of definition of the $X-Y$ plane, which must be open, connected and including the vacuum ($X=Y=0$). The condition $\varphi(X,Y)=\varphi(X,-Y)$ must be also satisfied in order to implement parity invariance. In addition we require the positive definiteness of the flat-space energy density

\begin{equation}
\rho=T_0^0=2X\varphi_X-\varphi(X) \geq \left(\sqrt{X^{2}+Y^{2}} + X\right) \frac{\partial \varphi}{\partial X}+ Y\frac{\partial \varphi}{\partial Y} - \varphi(X,Y) \geq 0,
\end{equation}
whose fulfillment implies the conditions $\varphi(0,0) = 0$, $\frac{\partial \varphi}{\partial X}\vert_{(X>0,Y=0)} > 0$ and $\varphi(X>0,0) > 0$ (see \cite{dr08}). These admissibility conditions endorse the definiteness and single-valued character of the ESS solutions and are essential in the analysis of the problem considered here. In particular, from the first-integral (\ref{FI}) it follows that $E(r)$ must be a monotonic function of $r$ for a given $q$. Moreover, for admissible models $\varepsilon_{ex}(r,q)$ is a \textbf{monotonically decreasing and concave} function of $r$, while $\varepsilon_{in}(r,q)$ is a \textbf{monotonically increasing and convex} one.

For finite-energy ESS solutions, the energy associated to the ESS field is given by

\begin{equation}
\varepsilon(q)=\int_0^{\infty} r^2 T_0^0(r,q)dr=q^{3/2}\varepsilon(q=1), \label{energy}
\end{equation}
where $\varepsilon(q=1)$ is the energy of the solution of unit charge, a universal constant for a given model. We can now split the families of admissible models into two sets, according to the character (finite or divergent) of the energy (\ref{energy}). Such a character only depends, for admissible models, on the behaviour of the ESS fields as $r \rightarrow \infty$ and at $r=0$. In this sense, we shall consider power-law expressions for the field behaviour around each of these limits. Let us then study each case separately.

\textbf{I)} \underline{Models with finite-energy ESS solutions}

Asymptotically the field must behave as

\begin{equation}
E(r)\sim r^p, \label{asymp}
\end{equation}
with $p<-1$ for convergence of the energy there. The behaviour of the associated lagrangian density is obtained, by making use of the first-integral (\ref{FI}), as

\begin{equation}
\varphi(X)\sim X^{\frac{p-2}{2p}}. \label{lagB}
\end{equation}
In what follows we shall call this asymptotic behaviour, given by Eq.(\ref{asymp}), as \textbf{B}-type fields. Such behaviour can be slower than Coulombian ($-2<p<-1$, subcase \textbf{B1}), Coulombian ($p=-2$, \textbf{B2}) or faster than Coulombian ($p<-2$, \textbf{B3}).

As $r \sim 0$ there are two ESS field behaviours compatible with the finite energy requirement (\textbf{A}-type fields) for admissible models. In case \textbf{A1} the field behaves at the center as in Eq.(\ref{asymp}) but now with $-1<-p<0$ for the convergence of the energy there (and, consequently, the field diverges in this case), and the lagrangian behaves also as in Eq.(\ref{lagB}). On the other hand, in case \textbf{A2} the field becomes finite at the origin, behaving there as

\begin{equation}
E(r) \sim a-br^{\sigma},
\end{equation}
where $a$ and $\sigma(>0)$ are fixed parameters for a given model while $b$ is a function of the charge \cite{dr09}. Concerning the lagrangian density, its behaviour around the central values of the fields is given by (the case $\sigma=2$ is singular, see Ref.\cite{dr08})

\begin{equation}
\varphi(X,Y=0)\sim \frac{2q\sigma b^{2/\sigma}}{2-\sigma}(a-\sqrt{X})^{\frac{\sigma-2}{\sigma}} + \Delta,
\end{equation}
where $\Delta=\varphi(a^2,Y=0)$ is an integration constant. Consequently we see that there are two different behaviours for $\varphi(X)$: if $\sigma \leq 2$ the lagrangian density diverges at $X=a^2$ while if $\sigma>2$ it takes a finite value there given by $\Delta$.

Consequently, for any admissible model $\varphi(X,Y=0)$ to support finite-energy ESS solutions, these fields must belong to one of the \textbf{B}-cases at $r \rightarrow \infty$, and to one of the \textbf{A}-cases around $r \sim 0$. In between both domains the lagrangian density must be a strict monotonically increasing function of $X$.

\begin{figure}
\begin{center}
\includegraphics[width=10cm,height=6.5cm]{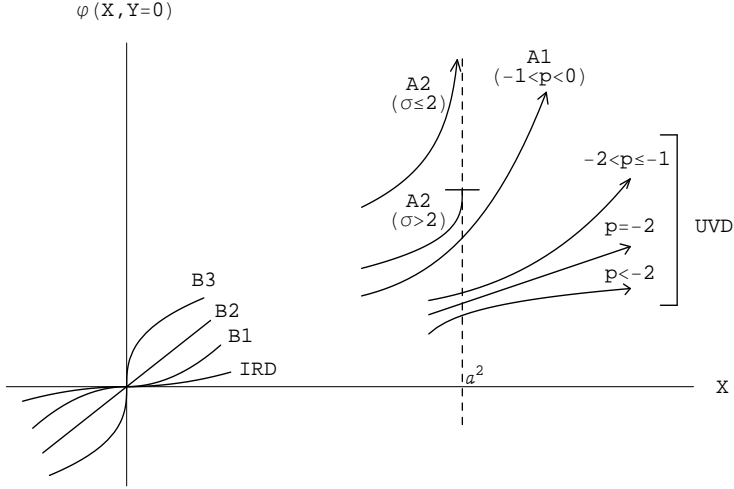}
\caption{\label{fig:1} Qualitative behaviour of the families of admissible lagrangian density functions $\varphi(X,Y=0)$. The behaviour at $X \sim 0$ ($r \rightarrow \infty$) is split into finite-energy cases (\textbf{B1}, \textbf{B2} and \textbf{B3}) and the energy-divergent one (\textbf{IRD}). At the center of the ESS solutions, there are two \textbf{A2} cases ($X \sim a^2$): for $\sigma \leq 2$, $\varphi(X)$ diverges there while if $\sigma>2$, $\varphi(X)$ becomes finite. \textbf{A1} is a vertical parabolic branch with divergent slope for large $X$. \textbf{UVD} is the family of divergent-energy solutions there, with its different possible slopes plotted.}
\end{center}
\end{figure}

\textbf{II)} \underline{Models with divergent-energy ESS solutions}

ESS fields whose energy diverges at $r=0$ are obtained from Eq.(\ref{asymp}), but now imposing the restriction $p<-1$ there. We shall call this kind of behaviour as \textbf{UVD} (meaning ``ultraviolet"-divergent). A similar behaviour is assumed in the asymptotic limit, with $-1\leq p<0$ for (asymptotically vanishing) ESS solutions with divergent energy there. The acronym \textbf{IRD} (``infrared"-divergent) will be employed for calling this family. Consequently the class of admissible NED models with divergent-energy ESS solutions is naturally classified into three families, resulting from the combination between i) divergent-energy fields at the center but not asymptotically (\textbf{UVD-B}), ii) asymptotic divergent-energy but not at the center (\textbf{A-IRD}) and iii) divergent energy in both limits (\textbf{UVD-IRD}). As a consequence of these behaviours $\varepsilon_{ex}(r,q)$ can only be defined in the first one of these cases, and $\varepsilon_{in}(r,q)$ only in the second one.

\subsection{Metric structure. Horizons}

With these ingredients we resume the previous discussion on the metric structure. First we look for the horizons $r_h$ of the different configurations, which are the solutions of $\lambda(r)=0$ in Eq.(\ref{metric1}) (in such cases where $\varepsilon_{ex}(r,q)$ can be defined, see the previous section), implying

\begin{equation}
M-\frac{r}{2}=\varepsilon_{ex}(r,q),
\end{equation}
and thus horizons are given by the cutting points between the monotonically decreasing and concave curve $\varepsilon_{ex}(r,q)$ and the beam of straight lines $M-\frac{r}{2}$, corresponding to different values of the ADM mass $M$ (see Fig.2). Let us now study, following this procedure, each class of models, resulting from the combination of one of the possible field behaviours at the center (\textbf{UVD, A1} or \textbf{A2}) with one of the asymptotic cases (\textbf{B1, B2, B3} or \textbf{IRD}).

\textbf{I) \underline{A1-B}:} At $r \rightarrow \infty$ the metric (\ref{metric1}) vanishes as ($p<-1$)

\begin{equation}
\lambda(r) \rightarrow 1-\frac{2M}{r}+\frac{32\pi q}{(p-2)(p+1)}r^p +\cdot \cdot \cdot \label{met-asym}
\end{equation}
and, consequently, the large-$r$ behaviour is governed by the Schwarzschild term, regardless of the \textbf{B}-subcase chosen (\textbf{B1}, \textbf{B2} or \textbf{B3}). At the center $\varepsilon_{ex}(r,q)$ takes the value $\varepsilon(q)$, with a negative divergent slope there. Consequently, the cutting points with the beam $M-\frac{r}{2}$ give the following configurations (see Fig.2): i) There is a family of extreme black holes (EBH), satisfying the conditions $\lambda(r)=\frac{d\lambda(r)}{dr} \big\vert_{r_h}=0$, which give the EBH radius and mass as

\begin{equation}
8\pi r_{hextr}^2T_0^0(r_{hextr},q)=1 \hspace{0.1cm};\hspace{0.1cm} M_{extr}(q)=\frac{r_{hextr}}{3}+\frac{16\pi q}{3} A_0(r_{hextr},q), \label{ebh}
\end{equation}
where $A_0(r,q)$ is the electric potential; ii) $M<M_{hextr}(q)$: naked singularity (NS); iii) $M_{hextr}(q)<M<\varepsilon(q)$: two-horizons (Cauchy and event) BH; iv) $M>\varepsilon(q)$: single-horizon BH; v) $M=\varepsilon(q)$: for this critical configuration we must study the behaviour of the metric around $r \sim 0$ ($-1<p<0$):

\begin{equation}
\lambda(r) \sim 1-\frac{2(M-\varepsilon(q))}{r}+ \frac{32\pi q}{(p-2)(p+1)}r^p + \cdot \cdot \cdot, \label{metrica1}
\end{equation}
and we see that in this case $\lambda(r)$ diverges there to $-\infty$, as in subcase iv) and, consequently, this is a single-horizon BH as well. In subcases iv) and v) the central singularity is spacelike, as opposed to the timelike singularity of subcases i), ii), and iii). \emph{Example of this family}: The Euler-Heisenberg lagrangian $\varphi(X,Y)=\frac{X}{2}+\mu(X^2+\frac{7}{4}Y^2)$, $\mu$ being a constant (see Ref.\cite{dr09} for details of this family in gravitation).

\begin{figure}[]
\begin{center}
\includegraphics[width=10cm,height=6.5cm]{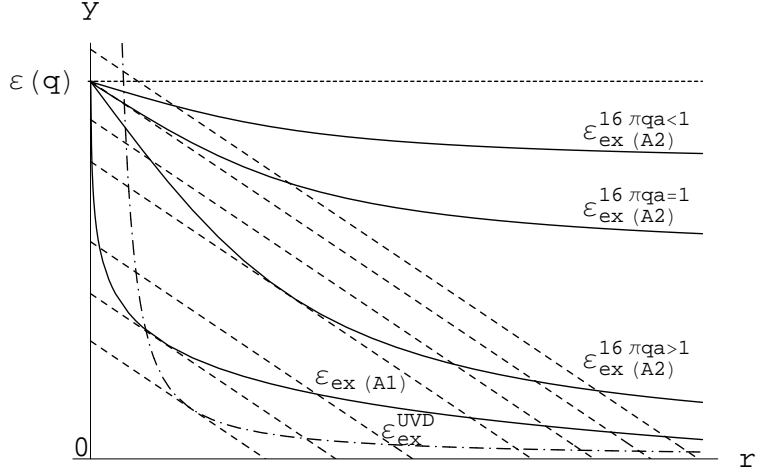}
\caption{\label{fig:2} Qualitative behaviour of $\varepsilon_{ex}(r,q)$ for the families \textbf{A1}, \textbf{A2} (three subcases) and \textbf{UVD} at the center combined with \textbf{B}-fields at $r \rightarrow \infty$, cut by a set of dashed lines, corresponding to different values of $M$. The cut points give the horizons of the different configurations (see the discussion in the text). All the finite-energy cases have been chosen in order to begin at the same value of $\varepsilon(q)$ as $r \rightarrow 0$.}
\end{center}
\end{figure}

\textbf{II) \underline{A2-B}:} Now there are three subcases, according to the (now finite) slope of $\varepsilon_{ex}(r,q)$ around $r \sim 0$. \textbf{A2a} ($16\pi qa>1$): Here there are five classes of configurations: i) $M<M_{hextr}(q)$: NS; ii) $M=M_{hextr}(q)$: EBH; iii) $M_{hextr}(q)<M<\varepsilon(q)$: two-horizons BH; iv) $M>\varepsilon(q)$: single-horizon BH; v) $M=\varepsilon(q)$: for this configuration the metric at the center is finite, behaving there as ($\sigma>0$)

\begin{equation}
\lambda(r) \sim 1-16\pi qa -\frac{2(M-\varepsilon(q))}{r}+ \frac{32\pi bq}{(\sigma +1)(2-\sigma)}r^{\sigma}+\Delta r^2,  \label{metrica2}
\end{equation}
and, consequently, in this case $\lambda(0)=1-16\pi qa<0$, leading to a single-horizon BH as well. \textbf{A2b} ($16\pi qa<1$): in this branch there are not EBH, and the available configurations are i) $M<\varepsilon(q)$: NS; ii) $M>\varepsilon(q)$: single-horizon BH; iii) $M=\varepsilon(q)$: in this case $\lambda(0)=1-16\pi qa>0$, as can be seen from (\ref{metrica2}), leading to a NS. \textbf{A2c} ($16 \pi qa=1$): the structures now are similar as in case \textbf{A1-B}, excepting for $M=\varepsilon(q)$, since $\lambda(0)=0$ and this configuration is a kind of ``black point" \cite{dr09}. \emph{Example of this family}: The BI lagrangian defined by Eq.(\ref{BI}).

\textbf{III) \underline{UVD-B}:} In this case $\varepsilon_{ex}(r,q)$ diverges as $r \rightarrow 0$ (since now $\varepsilon(q)$ becomes divergent, as explained in subsection 2.1.) and there are three possible configurations: i) $M<M_{hextr}(q)$: NS; ii) $M=M_{hextr}(q)$: EBH; iii) $M>M_{hextr}(q)$: two-horizons BH. \emph{Example of this family}: The Maxwell lagrangian $\varphi(X,Y=0)=\alpha X$ ($\alpha$ a constant), leading to the Reissner-Nordstr\"om solution $\lambda(r)=1-\frac{2M}{r}+\frac{q^2}{\alpha r^2}$.

\textbf{IV) \underline{A1 (and A2)-IRD}:} As already discussed, in this case $\varepsilon_{ex}(r,q)$ cannot be defined and the metric is integrated as in Eq.(\ref{metric2}). Now horizons are determined by the equation

\begin{equation}
\frac{r+C}{2}=\varepsilon_{in}(r,q), \label{cut-in}
\end{equation}
and by taking into account the monotonically increasing and convex character of $\varepsilon_{in}(r,q)$ the different configurations can be classified following the same procedure as in the previous cases, now using the cutting points with the beam of straight lines $\frac{r+C}{2}$ in Eq.(\ref{cut-in}), corresponding to different values of the integration constant $C$ \cite{dr10}. For the \textbf{A1} and \textbf{A2a} cases at the center we find: i) A family of EBHs, whose radius is given by that of Eq.(\ref{ebh}), now with an associated value $C=C_{extr}(q)$; ii) $C>C_{extr}(q)$: NS; iii) $0<C<C_{extr}(q)$: two-horizons BH; iv): $C<0$: single-horizon BH; v) $C=0$: single-horizon BH (now the metric diverges to $-\infty$ in case \textbf{A1}, while takes a finite value $\lambda(0)=1-16\pi qa<0$ in case \textbf{A2a}). In cases \textbf{A2b} and \textbf{A2c} the extreme black hole and two-horizons BH configurations disappear and we are lead to i) $C>0$: NS; ii) $C<0$: single-horizon BH; iii) $C=0$ a NS in case \textbf{A2b} and an extreme black point \cite{dr10} in case \textbf{A2c}. This class of gravitational configurations approaches asymptotic flatness slower than the Schwarzschild solution, since the asymptotic behaviour is governed by the ESS term (see Eq.(\ref{met-asym})). \emph{Example of this family}: The lagrangian $\varphi(X,Y=0)=X^{\gamma}$ with $3/2<\gamma<\infty$ (A1 family at the center).

\textbf{V) \underline{UVD-IRD}:} In this case neither $\varepsilon_{ex}(r,q)$ nor $\varepsilon_{in}(r,q)$ exist, but the above procedure for classifying the different gravitational structures works again, employing now Eqs.(\ref{metric3}) and (\ref{epsilon3}), together with the value of the integration constant $D$. This leads to: i) as in the previous cases, EBH configurations are found for a value $D=D_{hextr}(q)$; ii) $D>D_{hextr}(q)$: NS; iii) $D<D_{hextr}(q)$: two-horizons BH. \emph{Example of this family:} The lagrangian density $\varphi(X,Y=0)=X^{3/2}$.

\subsection{Extension to non-abelian fields}

The above results can be extended to generalized non-abelian gauge groups. Let us summarize the main ideas. The whole analysis will appear in a forthcoming publication. In this case the field strength components in the Lie algebra are defined as $F_{a\mu\nu}=\partial_{\mu}A_{a\nu}-\partial_{\nu}A_{a\mu}-g\sum_{bc}C_{abc}A_{b\mu}A_{c\nu}$, where $1\leq a \leq N$ ($N$ the dimension of the group) and $C_{abc}$ are the structure constants. The lagrangian density is again defined as $\varphi_{N-A}(X,Y)$ but now the field invariants are defined as (we are using the ordinary prescription in the calculation of the traces over the gauge group indices)

\begin{equation}
X=-\frac{1}{2}\sum_a (F_{a\mu\nu}F^{\mu\nu}_a)=\sum_a (\vec{E}^2_a-\vec{H}^2_a) \hspace{0.1cm}; \hspace{0.1cm} Y=-\frac{1}{2}\sum_a (F_{a\mu\nu}F^{*\mu\nu}_a)=2 \sum_a (\vec{E}_a \cdot \vec{H}_a).
\end{equation}
The associated field equations in presence of gravitation take the form

\begin{equation}
\sum_c \nabla_{ac\mu}\left[\varphi_X F^{\mu\nu}_c +\varphi_Y F^{*\mu\nu}_c \right]=0, \label{non-abelian}
\end{equation}
where $\nabla_{ac\mu}\equiv \delta_{ac}\nabla_{\mu}-g\sum_{bc}C_{abc}A_{b\mu}$ is the gauge-covariant derivative. Concerning the admissibility conditions, it can be shown that they take in the present case the same form as those obtained for NED models. For ESS solutions in this non-abelian case ($E_a \neq 0, H_a=0, \forall a$), the $N$ field equations arising from the $\nu=0$ component of the field equations (\ref{non-abelian}), in the metric (\ref{metric}), can be combined to give

\begin{equation}
r^2 \sqrt{X}\varphi_X \vert_{\footnotesize Y=0}=Q,
\end{equation}
where $Q=\sqrt{\sum_a^{N}Q_a^2}$ is the ``mean-square charge", constructed with the integration constants $Q_a$ arising from each one of the $\nu=0$ components of the field equations and identified as ``source color charges". As a consequence, there is a relation one-to-one between the ESS solutions of the generalized non-abelian gauge field theory and those of the NED models with the identification $\varphi_{N-A}(X,Y=0)=\varphi_{NED}(X,Y=0)$ and $q \leftrightarrow Q$. The associated energy $\varepsilon(Q)$ is the same as in the abelian case, but now there is a degeneration of energy in the sphere of radius $Q$. As a consequence of these equivalences, the integration of the metric from the Einstein equations, and the associated gravitational configurations for generalized non-abelian gauge groups can be shown to be the same as for the NED models analyzed above, once the previous identifications of lagrangian density functions and charges between abelian and non-abelian cases are performed.

\section{Conclusions and perspectives}

The gravitational configurations associated to general NED models minimally coupled to Einstein gravity can be qualitatively characterized without explicitly fixing the form of the lagrangian density function. In such a characterization the admissibility conditions play an essential role. A natural classification of the ESS solutions into several families is performed according to their asymptotic and central-field behaviours, which determine the character (finite or divergent) of the flat-space energy. Such NED models have been shown to lead to gravitating structures which are asymptotically Schwarzschild-like, or with an anomalous asymptotic non Schwarzschild-like behaviour (although still asymptotically flat and well behaved). In both cases the structure of horizons was studied, and explicit examples of lagrangian densities belonging to the different families of models were given. These procedures are not restricted to the asymptotically flat solutions of the Einstein gravity, but can be extended to (Anti-)de Sitter spaces and to higher-order curvature gravity theories (e.g. Gauss-Bonnet), a work in progress.

\ack{This work was partially supported by Spanish grants FC-08-IB08-154 and MICINN-09-FPA2009-11061. Hospitality of the organizers of the NEB14 meeting is acknowledged.}

\end{document}